\documentclass[12pt, a4paper]{article}
\usepackage{amsmath,amssymb,amsfonts}
\usepackage{graphicx}
\usepackage[colorlinks]{hyperref}
\usepackage{makecell}
\setcellgapes{3pt}

\begin{document}

\title{Induced vacuum current and magnetic flux in quantum scalar matter in the background of a  vortex defect with the Neumann boundary condition}

 \author{V.M. Gorkavenko${}^{1}$,  T.V. Gorkavenko${}^{1}$, Yu.A. Sitenko${}^{2}$, M.S. Tsarenkova${}^{1}$\\
  \it \small ${}^{1}$Department of Physics, Taras Shevchenko National
  University of Kyiv,\\ \it \small 64 Volodymyrs'ka str., Kyiv
  01601, Ukraine\\
  \it \small ${}^{2}$Bogolyubov Institute for Theoretical Physics,
  \it \small National Academy of Sciences of Ukraine,\\
  \it \small 14-b Metrologichna str., Kyiv 03680,
  Ukraine}
 \date{}

\maketitle

\begin{abstract}
A topological defect in the form of the Abrikosov-Nielsen-Olesen
vortex in a space of arbitrary dimension is considered as a gauge-flux-carrying tube that is
impenetrable for quantum matter. Charged scalar matter field is
quantized in the vortex background with the perfectly rigid
(Neumann) boundary condition imposed at the side surface  of the
vortex. We show that a current circulating around the vortex is induced in the vacuum,
if the Compton wavelength of the matter field exceeds
the transverse size of the vortex considerably. The vacuum current is periodic in the value of the gauge flux of the vortex,
providing a quantum-field-theoretical manifestation of the
Aharonov-Bohm effect. The vacuum current  leads to the appearance of an induced vacuum magnetic flux that for some values of the tube thickness exceeds the vacuum magnetic flux induced by a singular vortex filament. The results were compared to the results obtained earlier in the case of the perfectly reflecting
(Dirichlet) boundary condition imposed at the side surface  of the vortex.
It is shown that the absolute value  of the induced vacuum current and the  induced vacuum magnetic flux in the case of the Neumann boundary condition is  greater than in the case
of the Dirichlet boundary condition.

\end{abstract}

\maketitle

\section{Introduction}

There are many theoretical models in field theory which contain the phenomenon of spontaneous breakdown of  symmetries, see e.g. \cite{SBS}.
This phenomenon  gives rise to
topological defects  of various kinds,
see e.g. \cite{vilenkin,Brand}. In this paper, we will consider a linear topological
defect known as the Abrikosov-Nielsen-Olesen (ANO) vortex in condensed matter physics
\cite{Abr,NO}  or a cosmic string in  cosmology \cite{vilenkin,hind,Cop}. This topological object is formed  when the first homotopy group of the group space of the
broken symmetry group is nontrivial.
Abrikosov vortices  are  real physical objects in the type-II superconductors \cite{Hue,Rosenstein}.
Cosmic strings are currently hypothetical objects, and their possible manifestations such as gravitational waves, high-energy cosmic rays and gamma-ray bursts are actively searched for in the Universe \cite{Ber,Bra,Jack}.

In  classical theory of the ANO vortex, a spin-0 (Higgs) field  condenses and a spin-1 field
corresponds to the spontaneously broken gauge group; they are coupled  in the minimal way with constant
${\tilde e}_H$.
The transverse size of the vortex is of the order of
the correlation length or the  Compton wavelength, $\hbar(m_H c)^{-1}$, where $m_H$ is the mass
of the condensate field.  The physical requirements of
single-valuedness of the condensate field
and finiteness of the vortex energy result in the following dependence of the
vortex flux on  ${\tilde e}_H$: $\Phi=\oint d \textbf{x}  \textbf{A}
(\textbf{x})=2\pi \hbar c {\tilde e}_H^{-1}$, where $ \textbf{A}
(\textbf{x})$ is the vector potential of the  gauge field, and
the integral is over a path enclosing the vortex once. The
quantized matter field  is coupled minimally to the gauge field with
constant $\tilde e$. So quantum effects in the background of the
ANO vortex depend on the value of $\tilde e \Phi$.

Since
the phase with broken symmetry exists only outside the vortex, the quantum matter field
cannot  penetrate inside the vortex.
We   assume further that the
interaction between the ANO vortex and the quantized  matter field is
mediated   by the vector potential of the vortex-forming spin-1 field
only,
and the direct coupling between the vortex-forming spin-0 field
and the quantized  matter field can be neglected.
If so, the ANO vortex does not affect the
surrounding matter  in the framework of classical theory, and such an
influence is of purely quantum nature. The effect is a quantum-field-theoretical manifestation of the famous
Aharonov-Bohm effect \cite{Aha}, see review \cite{Pesh}. It is
characterized by the periodic dependence on the value of the vortex
flux, $\Phi$, with the period equal to the London flux quantum,
$2\pi \hbar c {\tilde e}^{-1}$.
A particular case of
${\tilde e}_H=2\tilde e$ $(\Phi=\pi \hbar c {\tilde e}^{-1}$, half
of the London flux quantum) is implemented in ordinary superconductors, see e.g. \cite{Hue}. Cases of fractional
values of the London flux quantum 
are physically meaningful as well,
and can be implemented in chiral
superfluids, liquid crystals and quantum liquids, see
\cite{Nel,Vol}.

The physical condition of non-penetration of the matter field inside the vortex means the absence of the matter field current through the side surface of the vortex, namely $\left. j_r \right|_{r_0}=0$, where $r$ is a radial coordinate which is perpendicular  to the side surface and $r_0$ is the  radius of the vortex.  Hence, the ANO vortex can be considered as a magnetic tube of the finite transverse size.
In the case of the scalar matter field, the condition of non-penetrability can be satisfied with the use  of a family of boundary conditions of the Robin type
\begin{equation}\label{Robin}
    (\cos \theta\, \psi + \sin \theta\, r \partial_r \psi)|_{r_0} =0,
\end{equation}
where $\theta$ is some arbitrary parameter. Among all possible values of  parameter $\theta$, two values are prominent.  The case of $\theta=0$ corresponds to the perfectly reflecting
(Dirichlet) boundary condition $\psi |_{r_0} =0$.
The case of $\theta=\pi /2$ corresponds the perfectly rigid
(Neumann) boundary condition $r \partial_r \psi|_{r_0}=0$.

In the present paper, we shall study the current  which is  induced in the vacuum of the quantized charged
scalar matter field by the ANO vortex with nonvanishing transverse size with  the perfectly rigid
(Neumann) boundary condition on its side surface ($\theta=\pi/2$). This current  creates a magnetic field in the vacuum, and the total induced vacuum magnetic flux will be studied in detail in this paper.
In the case of the ANO vortex of zero transverse size this problem was solved previously, see \cite{SitB13,SitB61,Sit515} and references
therein.
For the case of the finite transverse size of the ANO vortex  the induced vacuum current, magnetic flux, energy, and the Casimir force were considered in the \cite{Ivanch,newstring,newstring2,newstring3,newstring4} for the case of the perfectly reflecting
(Dirichlet) boundary condition.

It should be noted that for the quantized fermion matter field the condition of non-penetration of the matter field inside the finite transverse size ANO vortex has a form different from \eqref{Robin}, and it can be parameterized with the help of one parameter in the case of two-dimensional space and four parameters in the case of three-dimensional space.
The induced vacuum current and magnetic flux in these cases were considered for all values of  parameters  in \cite{FluxSpinorLow,FluxSpinor,FluxSpinor3}.

\section{Induced vacuum current and total magnetic flux}

We start with Lagrangian for a complex scalar field $\psi$  in
$(d+1)$-dimensional space-time
\begin{equation}\label{0}
\mathcal{L}=({\mbox{ $\nabla$}}_\mu\psi)^*({\mbox{
$\nabla$}}^\mu\psi)-m^2\psi^*\psi,
\end{equation}
where ${\mbox{ $\nabla$}}_\mu$ is the covariant derivative and $m$
is the mass of the scalar field. The vacuum current is conventionally defined as
\begin{equation}\label{a9}
\textbf{j}(\textbf{x})=-{\rm i}  \sum\hspace{-1.4em}\int\limits_{\lambda} (2E_\lambda)^{-1} \left\{ \psi_\lambda^*(\textbf{x})[{\mbox{\boldmath $\nabla$}} \psi_\lambda(\textbf{x})]-
[{\mbox{\boldmath $\nabla$}} \psi_\lambda(\textbf{x})]^*\psi_\lambda(\textbf{x})\right\},
\end{equation}
where $\lambda$ is the set of parameters (quantum numbers) specifying the
state, wave functions $\psi_\lambda(\textbf{x})$ form a complete set
of solutions to the stationary Klein-Fock-Gordon equation
\begin{equation}\label{3}
 \left(-{\mbox{\boldmath $\nabla$}}^2  + m^2\right)  \psi_\lambda({\bf x})=E^2_\lambda\psi({\bf x}),
\end{equation}
 $E_\lambda=E_{-\lambda}>0$ is the energy of the state;
symbol $\sum\hspace{-1em}\int\limits_\lambda$ denotes summation over
discrete and integration (with a certain measure) over continuous
values of  $\lambda$.

In the present paper we are considering a static background in the
form of the cylindrically symmetric gauge flux tube of the finite
transverse size. The coordinate system is chosen in  such a way that
 the tube is along the $z$ axis.
  The tube in 3-dimensional space is obviously generalized to
 the $(d-2)$-tube in $d$-dimensional space by adding extra $d-3$
 dimensions as longitudinal ones.
 The covariant derivative is $\nabla_0=\partial_0$, $\mbox{\boldmath
$\nabla$}=\mbox{\boldmath $\partial$}-{\rm i} \tilde e\, {\bf V}$
with $\tilde e$ being the coupling constant of dimension
$m^{(3-d)/2}$ and the vector potential possessing only  one
nonvanishing component given by
\begin{equation}\label{4}
V_\varphi=\Phi/2\pi,
\end{equation}
outside the tube; here  $\Phi$ is the value of the gauge flux inside
the $(d-2)$-tube and $\varphi$ is the angle in  polar $(r,\varphi)$
coordinates on a plane which is transverse to the tube.   The
Neumann boundary condition at the side surface of the tube
$(r=r_0)$ is imposed on the scalar field:
\begin{equation}\label{5}
\left.\partial_r\psi_\lambda\right|_{r=r_0}=0,
\end{equation}
i.e.the surface of the flux tube is  a perfectly rigid boundary for the matter field.

The solution to \eqref{3} and \eqref{5} outside the impenetrable
 tube of  radius $r_0$ takes form
\begin{equation}\label{6}
\psi_{kn{\bf p}}({\bf x})=(2\pi)^{(1-d)/2}e^{{\rm i}\bf{p
x}_{d-2}}e^{{\rm i}n\varphi}\Omega_{|n- {\tilde e}
\Phi/2\pi|}(kr,kr_0),
\end{equation}
where
\begin{equation}\label{7}
\Omega_\rho(u,v)=\frac{Y^\prime_{\rho}(v)J_{\rho}(u)-J^\prime_{\rho}(v)Y_{\rho}(u)}{\left[{J^{\prime}_{\rho}}^2(v)+{Y^{\prime }_{\rho}}^2(v)\right]^{1/2}},
\end{equation}
and $0<k<\infty$, $-\infty<p^j<\infty$ ($j=\overline{1,d-2}$), $n\in
\mathbb{Z}$ ($\mathbb{Z}$ is the set of integer numbers),
 $J_\rho(u)$ and $Y_\rho(u)$ are the Bessel functions of order $\rho$ of the first and  second
 kinds, the prime near the function means derivative with respect to the function argument. Solutions \eqref{6} obey orthonormalization condition
\begin{equation}\label{8}
\int\limits_{r>r_0} d^{\,d}{\bf x}\, \psi_{kn{\bf p}}^*({\bf
x})\psi_{k'n'{\bf p}'}({\bf
x})= \frac{\delta(k-k')}{k}\,\delta_{n,n'}\,\delta^{d-2}(\bf{p}-\bf{p}').
\end{equation}

Using \eqref{a9} and \eqref{6} we get
$j_r={\bf j}_{d-2}=0$ and
\begin{equation}\label{10}
j_\varphi(r)\equiv x^1 j^2(\textbf{ x})-x^2
j^1(\textbf{x})=(2\pi)^{1-d} \int d^{d-2}p\int\limits_0^\infty dk\,k
({\bf p}^2+k^2+m^2)^{-1/2}S(kr,kr_0),
\end{equation}
where
\begin{equation}\label{10a}
S(u,v)=\sum_{n\in\mathbb{Z}}\left(n-\frac{ \tilde
e\Phi}{2\pi}\right)\Omega_{|n- \tilde e\Phi/2\pi|}^2(u,v).
\end{equation}
Due to the infinite range of the summation, the last expression is
periodic in flux $\Phi$ with a period equal to $2\pi{\tilde
e}^{-1}$, i.e. it depends on quantity
\begin{equation}\label{11}
    F=\frac{ \tilde e\Phi}{2\pi}-\left[\!\left[\frac{\tilde e\Phi}{2\pi}\right]\!\right],
\end{equation}
where $[[u]]$ is the integer part of  quantity $u$ (i.e. the integer
which is less than or equal to $u$).

Let us rewrite \eqref{10a} in the form
\begin{equation}\label{11a}
S(u,v)=S_0(u)+S_1(u,v),
\end{equation}
 where
\begin{equation}\label{12}
S_0(u)= \sum_{n=0}^\infty\left[\left(n+1\!-\!F\right)J_{n+1-F}^2(u)\!-\!
\left(n\!+\!F\right)J_{n+F}^2(u)\right]
\end{equation}
and
\begin{equation}\label{15}
S_1(u,v)=\sum_{n=0}^\infty\left[(n+1-F)\Lambda_{n+1-F}(u,v)\right. - \left.(n+F)\Lambda_{n+F}(u,v)\right],
\end{equation}
where
\begin{equation}\label{16}
\Lambda_\rho(u,v)= \frac{J^{\prime\,2}_{\rho}(v)\left[Y_\rho^2(u)-J_\rho^2(u)\right]-2J^\prime_\rho(v)J_\rho(u)Y^\prime_\rho(v)Y_\rho(u)}{Y^{\prime\,2}_{\rho}(v)+J^{\prime\,2}_{\rho}(v)}\,.
\end{equation}

Vacuum current $j_\varphi$ circulating around the $(d-2)$-tube leads
to the appearance of the vacuum magnetic field with strength
$B^{3...d}$ directed along the $(d-2)$-tube; this is a consequence
of the Maxwell equation
\begin{equation}\label{16a}
r\partial_r B^{3...d}_{(I)}({r})=- e j_\varphi({r}),
\end{equation}
where coupling constant $e$ differs in general from $\tilde e$. The
total flux of the induced vacuum magnetic field across a plane which
is orthogonal to the $(d-2)$-tube is defined as
\begin{equation}\label{16d}
  \Phi^{(I)}_d=2\pi\int_{r_0}^\infty dr\,r B^{3...d}_{(I)}({r})
\end{equation}
and is given by expression
\begin{equation}\label{16e}
  \Phi^{(I)}_d= e\pi\int\limits_{r_0}^\infty dr\,r j_\varphi(r)
  \left(1-\frac{r_0^2}{r^2}\right).
\end{equation}
Inserting $j_{\varphi}(r)$ \eqref{10} and changing the order of
integration over $r$ and ${\bf p}$, we get, see \cite{Ivanch},
\begin{equation}\label{a1}
  \Phi^{(I)}_d= e m^{d-3}\frac{(4\pi)^{(2-d)/2} }{2 \Gamma (d/2)}
  \int\limits_0^\infty \frac{du}{\sqrt{1+u^{2/(d-2)} }} 
  \mathcal{D}(mr_0\, \sqrt{1+u^{2/(d-2)}}),
\end{equation}
where $\Gamma(v)$ is the Euler gamma-function and
\begin{equation}\label{a2}
  \mathcal{D}(y)=\!\int\limits_y^{\infty}\!dx \left(1-\frac{y^2}{x^2}\right)\int\limits_0^\infty
  \frac{dz\,z}{\sqrt{z^2+x^2}}\,S\left(z,z\frac{y}{x}\right).
\end{equation}
It should be noted that function $ \mathcal{D}(y)$ \eqref{a2} is
immediately related to the total induced vacuum magnetic flux in the
$d=2$ case:
\begin{equation}\label{a3}
\Phi^{(I)}_2=\frac{ e}{2m}\,\mathcal{D}(mr_0).
\end{equation}
Since $S_1(u,0)=0$, one can obtain
\begin{equation}\label{a4}
\mathcal{D}(0)=\int\limits_0^\infty dx
\int\limits_0^\infty\frac{dz\,z}{\sqrt{z^2+x^2}}\,
S_0(z)= \frac13\,F(1-F)\left(F-\frac12\right),
\end{equation}
and the total induced vacuum magnetic flux in the $d=2$ case is
finite in the limit of a singular (i.e. infinitely thin) vortex
filament, $r_0\rightarrow0$ \cite{SitB13}:
\begin{equation}\label{a5}
\lim_{r_0\rightarrow0}\Phi^{(I)}_2=\frac{ e}{6m}\,
F(1-F)\left(F-\frac12\right).
\end{equation}

However, in the $d\geq 3$ cases the total induced magnetic flux becomes infinite in the limit of the infinitely thin vortex  filament \cite{SitB61}. In this case, the consideration of the vacuum polarization by the vortex
filament of the finite transverse size is especially actual.

\section{Numerical analysis  of the induced vacuum characteristics}

As one can see from the previous section, in order to find the induced vacuum magnetic flux in $d$-dimensional  space, we first need to find it in 2-dimensional space. Unfortunately, this task in the case of  the vortex
tube of  finite transverse size  can be solved only by numerical methods.

With this aim we  rewrite expression \eqref{10} in the $d=2$ case in
dimensionless form\vspace{-0.5em}
\begin{equation}\label{17}
\left.rj_\varphi(r)\right|_{d=2}=\frac1{2\pi}\int\limits_0^\infty dz\,z
\left[z^2+\left(\frac{m
r_0}{\lambda}\right)^2\right]^{-1/2}S(z,\lambda z),
\end{equation}
where $\lambda=r_0/r$  $(\lambda\in [0,1])$.

In the limit of a singular  filament ($r_0=0$) expression \eqref{17}  contains only $S_0$  \eqref{12}. Summation in \eqref{12} can be performed
analytically and \eqref{17}  is reduced to
the following form, see \cite{Ivanch},
\begin{multline}\label{s1}
\left.rj_\varphi^{sing}(r)\right|_{d=2}=\frac{\sin(F\pi)}{\pi^3}
\int\limits_{mr}^\infty
dw
\frac{w^2}{\sqrt{w^2-(mr)^2}} \times \\
\left\{\!w\!\left[K^2_{1-F}(w)\!-\!K^2_F(w)\right]\!+\!(2F\!-\!1)K_F(w)K_{1-F}(w)
\right\},
\end{multline}
where $K_\rho(u)$ is the Macdonald function of order $\rho$.   The total
induced vacuum magnetic flux in this case, see \eqref{a5}, attains
the maximal absolute value equal to $|e|/(72\sqrt{3}m)$ at $F=F_\pm
$, where \vspace{-0.5em}
\begin{equation}\label{Flux}
  F_\pm=\frac12\left(1 \pm \frac1{\sqrt{3}}  \right).
\end{equation}

Next, we will numerically compute  the induced vacuum
current in the $d=2$ case   at $F=F_+$, when the
integral in \eqref{17} is likely to be most distinct from zero (note that the current at
$F=F_-$ equals to that at
$F=F_+$ with the opposite sign).
So, for a vortex tube of nonvanishing radius, we have to compute
values of dimensionless quantity $r j_\varphi$ at different values
of $\lambda$.
To do this, we perform high-precision numerical
integration in \eqref{17} with the help of a technique developed
earlier in \cite{Ivanch,newstring,newstring2,newstring3,newstring4}. The results of computation can be approximated
by an interpolation function in the form
\begin{equation}\label{18}
\left.rj_\varphi(r)\right|_{d=2}=
\left[\frac{e^{-2x}}{\sqrt{x}}\right]\left[\frac{P_n(x-x_0)}{x^n}\right]
\frac{Q_k(x^2)}{R_k(x^2)},\quad  x>x_0,
\end{equation}
where $x=mr$, $x_0=mr_0$ and $P_j(y)$, $Q_j(y)$, $R_j(y)$  are
polynomials in $y$ of the $j$-th order with the $x_0$-dependent
coefficients. It turns out that, for the interpolation of data, the most suitable choice of
function \eqref{18} contains the polynomials with indices $n = 9$ and $k= 4$. The first factor in square brackets describes the
large distance behavior in the case of a zero-radius tube
(filament), the second factor in square brackets is an
asymptotics at small distances from the side surface of the tube,
the last factor describes the behavior at intermediate distances.
Since the vortex tube is  impenetrable, $rj_\varphi(r)$ \eqref{18}
vanishes at $x\leq x_0$.

\begin{figure*}[t]
\begin{center}
\includegraphics[width=165mm]{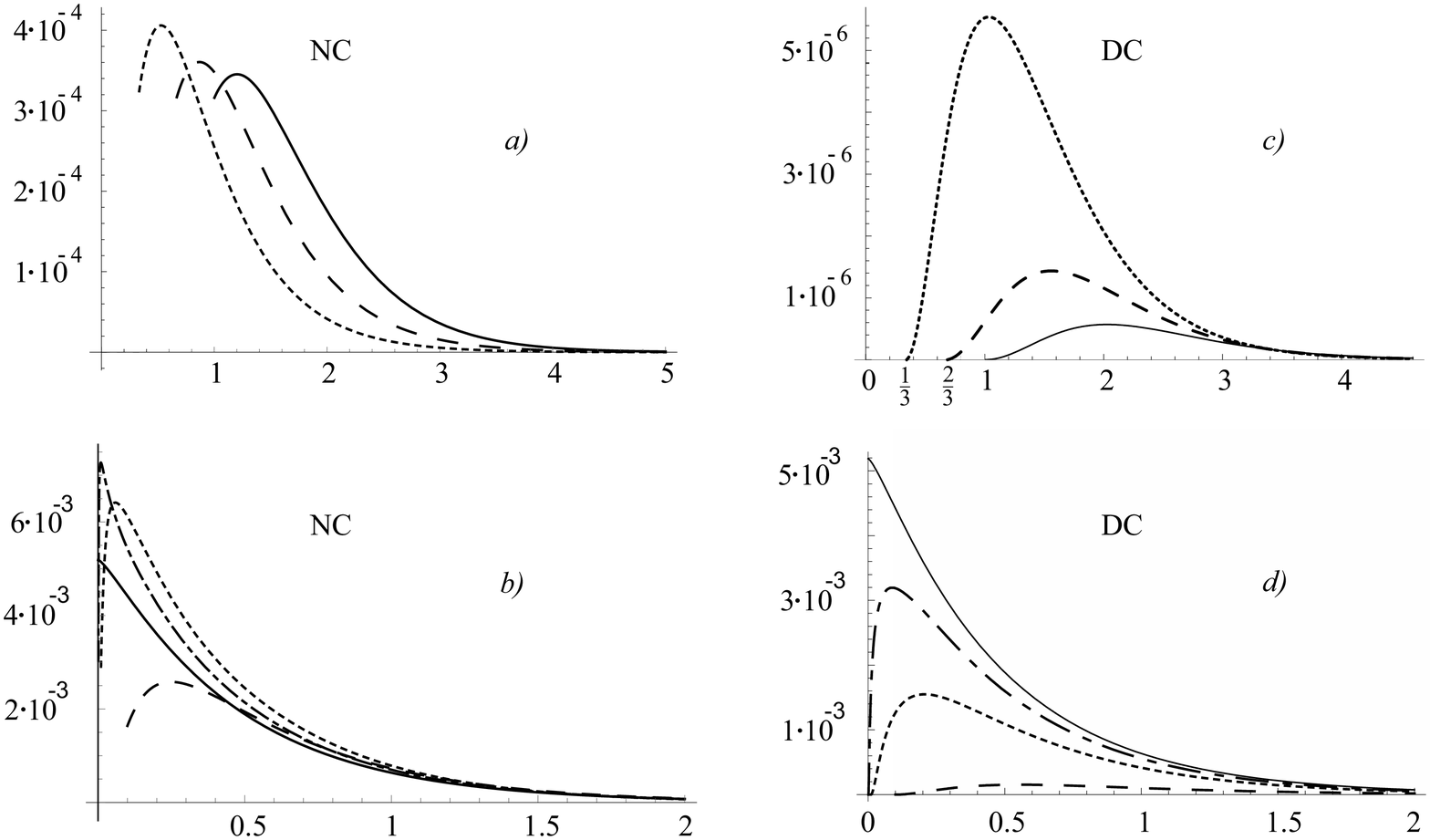}
\end{center}
\caption{The dimensionless induced vacuum current ($r j_{\varphi}$)
as a function of the dimensionless distance from the axis of the
tube $(x)$ for different values of the dimensionless tube radius
($x_0$) for the case of the Neumann (NC) and the Dirichlet (DC) boundary conditions:  (a), (c) solid line corresponds to $r j_{\varphi}\cdot 10^2$ for
$x_0=1$, dashed line corresponds to $r j_{\varphi}\cdot 10$ for
$x_0=2/3$ and dotted line corresponds to $r j_{\varphi}$ for
$x_0=1/3$; (b), (d)  solid line corresponds to the cases of a singular filament ($x_0=0$), dashed line corresponds to
$x_0=10^{-1}$, dotted line corresponds to
$x_0=10^{-2}$ and dash-dotted line corresponds to
$x_0=10^{-3}$. Variable $x$ is along the abscissa axis. \label{fig:1N}}
\end{figure*}

The results in the $d=2$ case for the induced vacuum current in the case of the Neumann boundary condition are presented on Fig.\ref{fig:1N}a and Fig.\ref{fig:1N}b.
For comparison, we also present the results for the induced vacuum current in the case of the Dirichlet  boundary condition \cite{Ivanch}, see Fig.\ref{fig:1N}c and Fig.\ref{fig:1N}d.
As one can see, the
current is negligible for the  tube of large radius, i.e. of order
of the Compton wavelength and greater, $r_0\geq m^{-1}$, see Fig.\ref{fig:1N}a and Fig.\ref{fig:1N}c. 
One can note that the induced vacuum current in the case of the Neumann boundary condition has a much weaker dependence on the
tube thickness $x_0$ (Fig.\ref{fig:1N}a) as compared to that in the case of
the Dirichlet boundary condition (Fig.\ref{fig:1N}c).
The current in the case of $r_0\ll m^{-1}$ is comparable with
the current in the case of a singular filament, see Fig.\ref{fig:1N}b and Fig.\ref{fig:1N}d. It should be noted that in the case of the Dirichlet boundary condition induced vacuum current
is always less in value than in the case of a singular filament, see Fig.\ref{fig:1N}d. It is not true for the case of the Neumann boundary condition, see Fig.\ref{fig:1N}b. It should be noted that the value  of the current in the case of the Neumann boundary condition is   greater than the value of the current in the case of the Dirichlet boundary condition at any value of the tube thickness.

Using \eqref{16e}, \eqref{a2}, \eqref{a3} and \eqref{18}, we compute numerically the total induced vacuum magnetic flux in the $d=2$
case for different values of the tube thickness (parameter $x_0=mr_0$). The results of the computation in the dimensionless form can be approximated
by an interpolation function in the form \vspace{-1em}
\begin{equation}\label{flux}
\ln \frac{m \Phi^{(I)}_2}{e}=\ln \mathcal{D}(x_0) = M_1(x_0) \Theta(X-x_0)+ \left[a + b x_0^1 + d x_0^5 + \sqrt{L_2(x_0)} \right] \Theta(x_0-X),
\end{equation}
where  $M_j(y)$, $L_j(y)$  are
polynomials in $y$ of the $j$-th order, $\Theta(y)$ is the Heaviside step function and $X=10^{-2}$.
Using \eqref{a1} and \eqref{flux},  we compute
numerically the total induced  magnetic flux in the $d=3,4$
cases, i.e. $\Phi_3^{(I)}$ and $\Phi_4^{(I)}$. The
results for the $d=2,3,4$
cases are presented on Fig.\ref{fig:flux}a and in Table \ref{Table1}. For comparison,  we also present  results for the induced  flux in the case of the Dirichlet boundary condition \cite{Ivanch}, see Fig.\ref{fig:flux}b and Fig.\ref{fig:flux}c.

As one can see, in the case of the Neumann boundary condition (for  $d=2$) there is a region of the tube  thickness $(0<x_0<0.4)$
where the absolute value of the induced flux is greater than
in the case of a singular filament, see Fig.\ref{fig:flux}a. Whereas
the absolute value of the flux induced by a singular filament  is always
greater than the absolute value of the flux induced by a tube of the
nonvanishing radius with the Dirichlet boundary condition, see Fig.\ref{fig:flux}b and Fig.\ref{fig:flux}c.
One can see also that the induced flux from a tube with the Neumann boundary condition is greater than the induced flux from a tube with the Dirichlet boundary condition at any value of the tube thickness.

In the case of higher space dimensions the behavior of the induced flux in the case of the Neumann and Dirichlet boundary conditions is similar, see Fig.\ref{fig:flux}.
For the case of a space with a higher dimension ($d=4$) at large radii of the tube the induced flux is a more strongly decreasing function, but at smaller radii of the tube the induced flux  is a more strongly increasing function compared to that in the case of the space with a smaller dimension ($d=3$).
While the induced flux in the
unphysical case of a singular filament is infinite for $d>2$, the induced flux in
the physical case of a tube of nonvanishing radius is finite. As one can see from Table \ref{Table1}, the absolute value of the flux induced by a tube with the Neumann boundary condition
is always greater than the absolute value of the flux induced by a tube with the Dirichlet boundary condition for any dimension of the space.

\begin{table*}
\centering\makegapedcells
\begin{tabular}{|c|c|c|c|c|c|c|}
\hline $x_0$ & 1 & 2/3 & $1/3$ & $10^{-1}$ & $10^{-2}$ & $10^{-3}$\\
\hline
$\dfrac{m }{e}\,\Phi^{(I)}_2$, NC & $7.09\cdot 10^{-6}$ &
$7.79\cdot10^{-5}$ & $9.05\cdot10^{-4}$
& $5.34\cdot 10^{-3}$ & $9.92\cdot 10^{-3}$ & $0.01$\\
\hline $\dfrac{m }{e}\,\Phi^{(I)}_2$, DC & $2.36\cdot 10^{-8}$ &
$5.76\cdot10^{-7}$ & $2.07\cdot10^{-5}$
& $4.88\cdot 10^{-4}$ & $3.66\cdot 10^{-3}$ & $6.22\cdot 10^{-3}$\\
\hline
\hline
$\dfrac{1}{e}\,{\Phi^{(I)}_3} $, NC & $ 10^{-6}$ &
$1.41\cdot10^{-5}$ & $2.26\cdot10^{-4}$
& $2.14\cdot 10^{-3}$ & $9.34\cdot 10^{-3}$ & $1.64\cdot 10^{-2}$\\
\hline
$\dfrac{1}{e}\,{\Phi^{(I)}_3} $, DC & $3.07\cdot 10^{-9}$ &
$8.83\cdot10^{-8}$ & $4.13\cdot10^{-6}$
& $1.48\cdot 10^{-4}$ & $2.22\cdot 10^{-3}$ & $6.55\cdot 10^{-3}$\\
\hline
\hline $\dfrac{1}{m e}\,{\Phi^{(I)}_4} $, NC & $7.5\cdot 10^{-8}$ &
$1.3\cdot10^{-6}$ & $3.02\cdot10^{-5}$
& $5.58\cdot 10^{-4}$ & $1.08\cdot 10^{-2}$ & $1.15\cdot 10^{-1}$\\
\hline
$\dfrac{1}{m e}\,{\Phi^{(I)}_4} $, DC & $2.06\cdot 10^{-10}$ &
$7.06\cdot10^{-9}$ & $4.43\cdot10^{-7}$
& $2.68\cdot 10^{-5}$ & $1.25\cdot 10^{-3}$ & $1.32\cdot 10^{-2}$\\
\hline
\end{tabular}
\vspace{1.5em}
\caption{The dimensionless induced vacuum magnetic flux in
cases of dimension $d=2,3,4$  for tubes of  different radii and for the case of the Neumann (NC) and Dirichlet (DC) boundary condition.}\label{Table1}
\end{table*}

\section{Summary}

In the present paper, we consider the current and the total magnetic flux
which are induced in the vacuum of the quantized charged scalar
matter field by a topological defect in the form of the ANO vortex in  a space of arbitrary dimension. We assume that the 

\begin{figure}[t]
\begin{center}
\includegraphics[width=\textwidth]{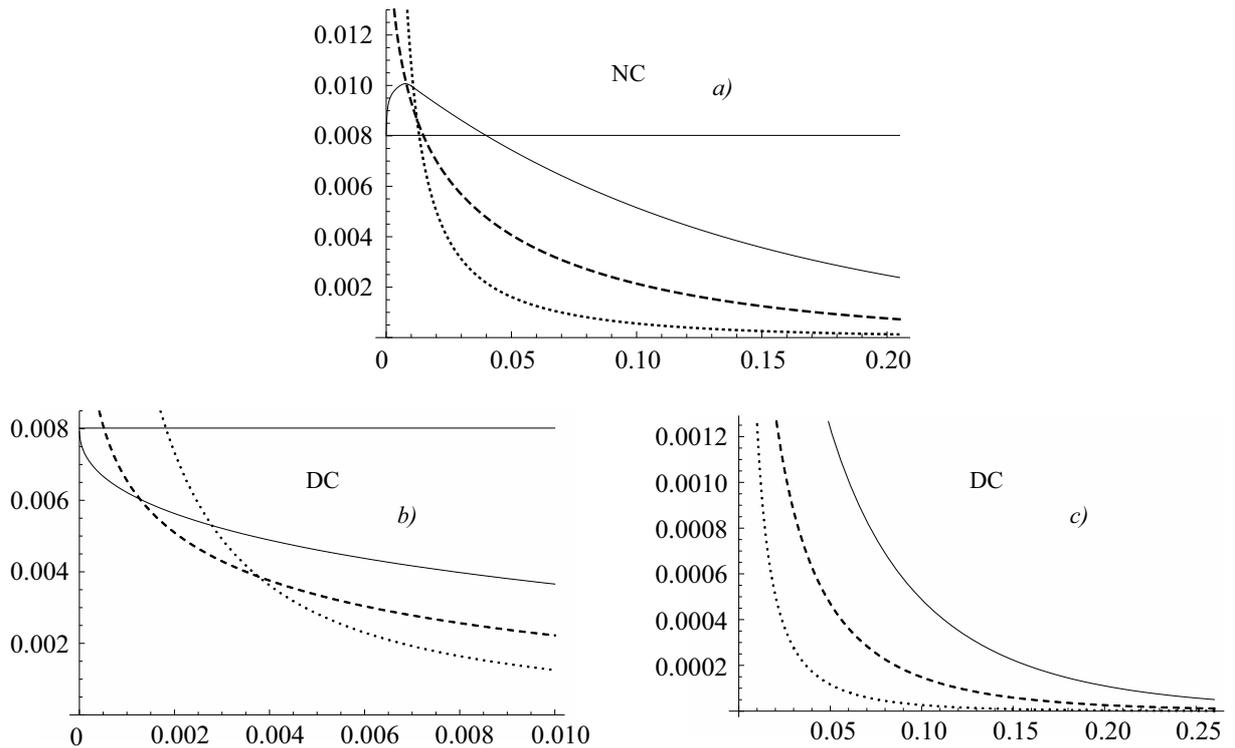}
\end{center}
\caption{The dimensionless  induced vacuum  magnetic flux in spaces
of different dimensionality as a function of the dimensionless tube
radius $(x_0)$ for the case of the Neumann (NC) and the Dirichlet (DC) boundary conditions:  $e^{-1}m \Phi^{(I)}_2$ -- solid line,
$e^{-1}\Phi^{(I)}_3$ -- dashed line,  $(em)^{-1}\Phi^{(I)}_4$ --
dotted line.   The case of $x_0=0$ and $d=2$ is
presented by a horizontal solid line. \label{fig:flux}}
\end{figure}

\noindent ANO vortex is impenetrable for quantum matter. The perfectly rigid (Neumann) boundary condition is imposed on
the matter field at the side surface of the vortex.
The same problem was considered previously for the vortex with the perfectly reflecting (Dirichlet) boundary condition on its side surface \cite{Ivanch}.
So, we compare results obtained for the Neumann and  Dirichlet boundary condition at the side surface of the vortex.

In both cases of the above mentioned boundary conditions, the induced
current is circulating around the vortex and it is vanishingly small in the case of the vortex
transverse size being of the order of or exceeding the Compton
wavelength of the matter field ($x_0 \gtrsim 1$), see Fig.\ref{fig:1N}, Fig.\ref{fig:flux} and Tabl.1 for the $d=2,3,4$ dimensions of the space.
It confirms  previously obtained  conclusion \cite{newstring,newstring2,newstring3,newstring4}  that
the vacuum polarization effects are essential only for matter fields with
masses which are much smaller than the scale of the spontaneous
symmetry breaking (mass of the Higgs  field forming the topological defect).

In both cases of the above mentioned boundary conditions,
the induced vacuum current decreases exponentially at large distances from the vortex. The  current and the induced vacuum magnetic  flux are odd
in the value of the vortex flux, $\Phi$, and periodic in this value
with  the period equal to the London flux quantum, $2\pi {\tilde
e}^{-1}$. They vanish at $F=0,1/2,1$  and are of opposite signs in the intervals
$0<F<1/2$ and $1/2<F<1$, with their absolute values being symmetric
with respect to the point $F=1/2$. In the case of space of dimension $d>2$, taking into account finite transverse size of the vortex
eliminates an unphysical divergence  for
the total induced vacuum  flux, which takes place in the case of a singular
vortex filament.

The visible difference between the cases of the Neumann and Dirichlet boundary conditions lies in the magnitude of the vacuum polarization effects.
The absolute value of the induced vacuum current and the  induced vacuum magnetic flux in the case of the Neumann boundary condition is greater than that in the case of the Dirichlet boundary condition.
In particular, as one can see from Fig.\ref{fig:1N}d and Fig.\ref{fig:flux}b for dimension of space $d=2$, the vacuum effects in the case of the  Dirichlet boundary condition are always  smaller than in the case of a singular vortex filament.
However, as one can see from Fig.\ref{fig:1N}b, for the case of the Neumann boundary condition, the absolute value of the induced vacuum current can exceed
the absolute value of the induced vacuum current in the case of a singular vortex filament. Moreover,
there is a region of  the vortex thickness values ($0<x_0<0.4$), see  Fig.\ref{fig:flux}a, where the absolute value of
the induced vacuum magnetic   flux  is greater  than that in the case of a singular vortex filament for the $d=2$ space.

\section{Acknowledgments}

The work was supported by the National Academy of Sciences of Ukraine (Project No.01172U000237).

\end{document}